\documentclass[lettersize,journal]{IEEEtran}
\usepackage{amsmath,amssymb,amsfonts}
\usepackage{amsthm}
\usepackage{algorithmic}
\usepackage{algorithm}
\usepackage{array}
\usepackage[caption=false,font=small,labelfont=sf,textfont=sf]{subfig}

\usepackage{textcomp}
\usepackage{stfloats}
\usepackage{url}
\usepackage{verbatim}
\usepackage{graphicx}
\usepackage{cite}
\usepackage{arydshln}
\usepackage{multirow}
\usepackage{xcolor}
\usepackage{threeparttable}
\usepackage[colorlinks, linkcolor=blue, citecolor=blue, urlcolor=blue]{hyperref}

\allowdisplaybreaks 
\begin{document}

\title{Real-time Operation of Electric Autonomous Mobility-on-Demand System Considering Power System Regulation}

\author{
Lyuzhu~Pan,~\IEEEmembership{Graduate Student Member,~IEEE,}
Hongcai~Zhang,~\IEEEmembership{Senior Member,~IEEE}
\thanks{L. Pan, and H. Zhang are with the State Key Laboratory of Internet of Things for Smart City and Department of Electrical and Computer Engineering, University of Macau, Macao, 999078 China (email: hczhang@um.edu.mo).}
}



\maketitle

\begin{abstract}
Electric autonomous mobility-on-demand (EAMoD) systems are emerging all over the world. However, their potential swarm charging in depots may deteriorate operation of the power system, further in turn affecting EAMoD system's optimal operation. To prevent this latent risk, we develop a real-time coordination framework for the EAMoD system and the power system. First, the temporal-spatial characteristics of EAMoD fleets are fully described based on a Markov decision process model, including serving trips, repositioning, and charging. Second, charger accessibility of EAMoD depot charging is well modeled as real-world configuration, wherein fast and slow charge piles are both included. Third, the power system regulation model provides real-time charging regulation constraints for EAMoD systems to prevent potential overload and undervoltage. To address the poor solution quality attributed to the complex decision space of the EAMoD system, this paper proposes a piecewise linear-based approximate dynamic programming algorithm combined with model predictive control. Numerical experiments in the Manhattan and a 14-node power distribution network validate the effectiveness of the proposed algorithm and underscore the necessity of system coordination.

\end{abstract}


\section{Introduction}
\IEEEPARstart{A}{utonomous} electric vehicles are widely regarded as a promising pathway to the future of urban transportation due to their enhanced efficiency and safety. As the vital stage in the evolution of automated transportation, \textit{electric autonomous mobility-on-demand} (EAMoD) systems, especially for ride-hailing service, have become increasingly mature worldwide over the past decade\cite{zhang2024sustainable}. For instance, Apollo Go has provided over 8 million trips and accumulated a test mileage exceeding 130 million kilometers by the end of 2024\cite{wheeler2025how}. However, large-scale EAMoD systems in the future can lead to the negative impact on the power system for two reasons. On the one hand, EAMoD fleets are centrally controlled by the dispatch center, and typically charge at centralized EAMoD depots for efficient management, such as operation paradigms of Waymo\cite{herger2024waymo} and Zoox\cite{herger2024zoox}. As a result, the charging power at depots will be significant. On the other hand, since there are no labor costs, EAMoD fleets are particularly sensitive to electricity costs, thus leading to the latent risk of the swarm charging. Consequently, EAMoD systems in distribution networks may cause various grid challenges, such as substation overload and undervoltage.


In recent years, a growing line of studies has been directed at the integration of autonomous vehicle systems within the power network. These works can be categorized into two groups in terms of transportation model types: the \textit{vehicle-routing model} and the \textit{flow-optimization model}. The vehicle-routing models have high model accuracy, where states and decisions of EAMoD fleets are represented by binary variables. For instance, Zhang \emph{et al. }\cite{zhang2016model} proposed a mixed-integer programming model, where a large number of binary variables were used to depict fleets' location, battery state, actions, etc. Building on this model, Iacobucci \emph{et al. }\cite{iacobucci2019optimizationa} and Yang \emph{et al. }\cite{yang2024optimal} further accounted for the power system model to investigate its impact on the mobility service quality. Melendez \emph{et al. }\cite{melendez2020optimal} proposed a comprehensive framework for coordinating autonomous electric fleets, energy hubs, and the power grid, incorporating both day-ahead and real-time stages. {However, the extensive use of binary variables results in poor computational efficiency, especially for the problem of large-scale coupled transportation and power system coordination. Moreover, compared to the classical vehicle-routing paradigm, additional charging (or energy) constraints in the EAMoD system will further deteriorate computational performance. In order to adapt to urban-scale vehicle scheduling scenarios, the flow-optimization model is widely accepted, which formulates the complex EAMoD decisions into continuous traffic flows, rather than integer variables.} For example, Rossi \emph{et al. }\cite{rossi2020interaction,estandia2021interaction} proposed a time-space-battery expanded network to investigate the mutual impact and equilibrium of the coupled EAMoD and power systems. Additionally, the network flow model has been applied to various problems, including the charging service pricing \cite{wang2024charging}, battery swapping scheduling \cite{ding2021integrated}, autonomous electric fleet sizing and charging infrastructure planning \cite{tian2024integrated,zhang2020joint}, and emission-oriented autonomous fleet dispatch \cite{sheng2023emissionconcerned}. {However, the above flow-optimization studies formulate the time inherently into the model, which is not suitable for the reformulation into dynamic form. Therefore, they typically ignore the temporal dynamics and the uncertainties in the traffic demand. As a result, the corresponding solution denotes an ideal result and fails to capture the real-time interactions between the EAMoD system and the power system. Besides, studies in both vehicle-routing and flow-optimization models generally account for only one charging type in the EAMoD system, which cannot reflect the practical configuration (i.e., equipped with both fast and slow charging facilities) in the EAMoD depot.}      

When the EAMoD dispatch model is coordinated with the power system over time, it becomes a sequential decision-making task. Three algorithms are commonly adopted to solve the sequential decision problems in EAMoD dispatch, including \textit{model predictive control} (MPC), \textit{reinforcement learning}, and \textit{approximate dynamic programming} (ADP). MPC algorithm optimizes the fleet decisions over a specified time horizon and executes the decision at the first time step. Kang \emph{et al. }\cite{kang2021maximumstability} and Chu \emph{et al. }\cite{chu2022joint} improved the traffic service quality via the real-time dispatch of autonomous fleets based on MPC algorithm. {However, MPC algorithm is limited by the myopic nature within the time horizon and its high computational burden. On the one hand, owing to the myopic nature, MPC algorithm might run out of all battery energy without charging, since the benefits of charging may occur out of the optimization horizon. On the other hand, MPC algorithm is equivalent to solve a problem with a scale several times larger than a single-period sub-problem.} Reinforcement learning algorithm learns the optimal decisions of EAMoD systems via repeated interaction with the environment. Shi \emph{et al. }\cite{shi2020operating}, Bagherinezhad \emph{et al. }\cite{bagherinezhad2023realtime} and Ahadi \emph{et al. }\cite{ahadi2023cooperative} employed deep Q-network based multi-agent reinforcement learning algorithms to determine the real-time routing and charging of EAMoD systems. Xie \emph{et al. }\cite{xie2023twosided} proposed a two-sided deep reinforcement learning algorithm based on the actor-critic algorithm to optimize the operation of the coupled autonomous and human-driven vehicles. {While reinforcement learning algorithm shows high computational efficiency in real-time applications, it has inherent limitations, including high data requirements, poor convergence, and high sensitivity to hyperparameters. Moreover, reinforcement learning is limited to a reasonable dimension of action space, which hinders its applications in high-dimensional scenarios.} ADP algorithm trains the value function based on the Lagrangian multipliers or marginal value of the model and then makes decisions according to the trained value function. Compared to model-free reinforcement learning, ADP algorithm requires accurate model information, allowing it to converge more stably and require less data\cite{powellapproximate}. Hu \cite{hu2025optimizing} investigated the dynamic coordinated dispatch of autonomous electric taxi and mobile charging vehicle using ADP, while Al-Kanj \emph{et al. }\cite{al-kanj2020approximate} applied a lookup table-based ADP for the real-time decision-making of autonomous fleets. The basic ADP algorithm can effectively handle the vehicle dispatch without charging decision\cite{schmid2012solving,chen2024realtime}, or under the assumption that autonomous fleets can charge anywhere\cite{al-kanj2020approximate}. \textcolor{black}{However, in practice, autonomous fleets usually can only charge at the specified EAMoD depots, which significantly increases the decision complexity, thus the basic ADP algorithm may fail to address this challenge (see further discussion in Section~\ref{subsec:adpmpc}).}   

\textcolor{black}{In this study, we propose a real-time EAMoD dispatch model under the power system regulation, which comprehensively accounts for both detailed temporal-spatial characteristics of EAMoD fleets, computational complexity at the urban scale, and both fast and slow charging modes. To address the challenges stemming from the complex decision space of EAMoD system when considering the charger accessibility into the model, we propose an improved ADP integrated with MPC.} The main contributions of our paper are twofold:
\begin{enumerate}
    \item We propose an integrated real-time operation model involving EAMoD dispatch and power system regulation based on the Markov decision process. The proposed model comprehensively incorporates the spatiotemporal dynamics of EAMoD fleets, uncertainties in trip requests, and power system regulation constraints. We further accurately model charger accessibility to mimic the real-world EAMoD depot charging, where depots are equipped with both fast and slow charging infrastructures. This model formulates the fleet behaviors in a flow-like manner, achieving more moderate computational complexity.
    \item We propose an ADP algorithm based on piecewise linear value function approximation to capture the future values of EAMoD states. The slope of the value function is updated according to the dual variable associated with the flow conservation constraint. Moreover, we combine the basic ADP with MPC to manage the challenge from the
 complex decision space, demonstrating computational superiority and near-optimal solutions compared to the basic ADP or MPC algorithm alone.
\end{enumerate}
\noindent In addition, numerical experiments are carried out with the real taxi data from Manhattan, New York. The result verifies the effectiveness of the proposed algorithm and underscores the necessity of considering power system coordination. 

The remainder of the paper is organized as follows. We present the problem formulation and develop the EAMoD dispatch and power system regulation models in Section \ref{problem statement}. We then introduce the proposed solution algorithm in Section \ref{solution method}. Numerical experiments are conducted in Section \ref{result}. The conclusions are provided in Section \ref{conclusion}.

\section{Problem Formulation}\label{problem statement}
\begin{figure}
    \centering
    \includegraphics[scale=0.5]{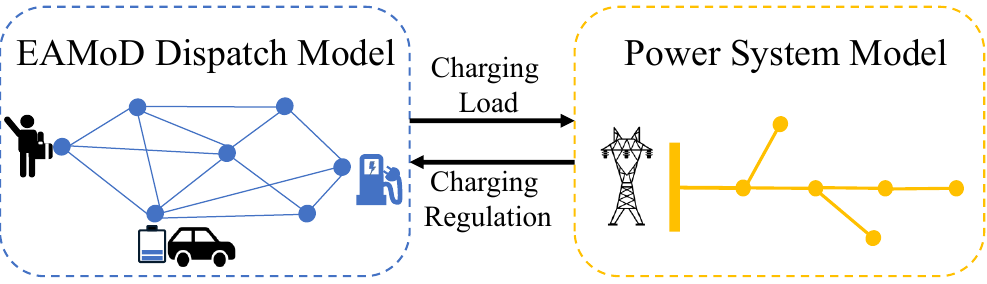}
    \caption{Interdependency between the EAMoD system and the power system.}
    \label{fig:coordination}
\end{figure}

The relationship between the EAMoD systems and the power system is shown in Fig.~\ref{fig:coordination}. \textcolor{black}{In this paper, the EAMoD system makes real-time decisions, including serving trips, repositioning to high-demand locations, and charging, to maximize fleet revenue. However, charging behaviors at EAMoD depots may deteriorate the economic and safe operation of the power system. Thus, the power system must stipulate the dynamic charging regulation constraints according to the base load prediction over time, and communicate the constraints to the EAMoD system. The communication can be implemented via the normal communication protocol since the charging regulations are aggregated values without confidential power network information. The EAMoD dispatch problem is a linear program, while the power system regulation problem is a second-order cone program. It should be noted that the EAMoD dispatch model is not suitable for the vehicle-level locational dispatch due to the solution continuity. Nevertheless, the continuous solutions are acceptable for the large-scale regional dispatch, where the vehicle scale and research area are large.} In this paper, the variable in the calligraphic letter denotes the set, while the bold variable denotes the vector or matrix. 

\subsection{Electric Autonomous Mobility-on-Demand Dispatch Model}\label{robotaix model}
In the model, the central operator dispatches the fleets to travel over time and space. Time is divided into the same intervals of 5 minutes. Although the decision-making process is discrete, trip requests are continuously received by the central operator, who dispatches EAMoD fleets at the end of each time interval. The unmet trip requests are assumed to be canceled. The city is divided into several zones $z\in\mathcal{Z}$, wherein $\mathcal{Z}^\text{Depot}\subset\mathcal{Z}$ denotes the zones containing depots. EAMoD fleets are assumed to be able to travel between zones without transportation accessibility restrictions, while travel time varies across zones. \textcolor{black}{Generally, the state of charge (SoC) is a continuous ratio to denote the remaining battery energy. In this paper, SoC is one of the attributes to categorize the vehicle state, which is an integer number. Therefore, we use SoC to denote the integer battery situation for presentation brevity afterwards. The battery capacities of all EAMoD vehicles are assumed to be the same, and the batteries are discretized into $|\mathcal{L}|$ levels, with $l\in\mathcal{L}$ as the SoC. The vehicle with SoC=1 (i.e., $l=1$) cannot move, while the vehicles with SoC=$|\mathcal{L}|$ are fully charged. In this study, the EAMoD dispatch model is a linear program with linear decisions.}

We model the EAMoD dispatch problem as a Markov decision process, including state, decision, transition function, and objective function, detailed as follows.

\subsubsection{State}
\begin{figure}
\centerline{\includegraphics[scale=0.7]{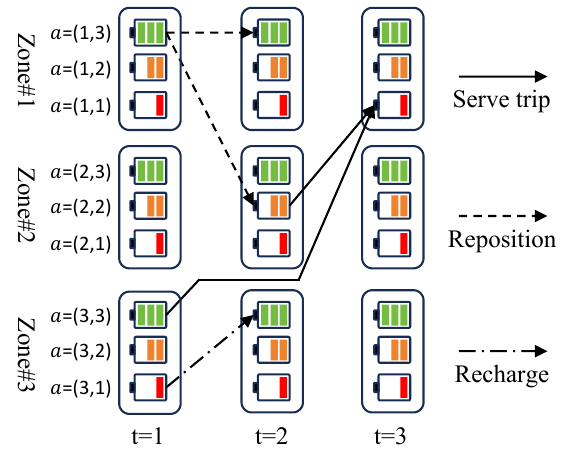}}
\caption{Illustration of EAMoD dispatch model. The simple model consists of three zones with three SoC levels.}
\label{fig:illustration}
\end{figure}

The state $\boldsymbol{S}_t$ of this system at time $t$ includes the vehicle state $\boldsymbol{R}_t$ and the demand state $\boldsymbol{D}_t$, i.e. $\boldsymbol{S}_t=(\boldsymbol{R},\boldsymbol{D})_t$, representing the number of vehicles and demand requests at time $t$, respectively. As the vehicle state $\boldsymbol{R}_t$ denotes the number of vehicles, we use the attribute $a$ as the index of the vehicle state to clarify the vehicle location and SoC:
\begin{align*}
    a=(z,l),
\end{align*}
where $z\in\mathcal{Z}$ denotes the location and $l\in\mathcal{L}$ denotes the SoC. Note that the attribute $a$ is a compact subscript, consisting of both location $z$ and SoC $l$. Using subscript $a$ is equivalent to subscript $(z,l)$. \textcolor{black}{With this modeling, we can estimate the scale of the state space according to $|\mathcal{A}|=|\mathcal{Z}|\times|\mathcal{L}|$.} Then, $R_{t,a}$ represents the vehicle number with specific attribute $a$ at time $t$. For instance, in the illustration (see Fig.~\ref{fig:illustration}), there are 9 vehicle attributes, including three locations and three SoC levels. Suppose there are 10 vehicles with attribute $a=(2,2)$ at time $t=1$, then the state is $R_{t=1,a=(2,2)}=10$. Similarly, we use $\boldsymbol{D}_t=[D_{t,b}]$ to represent the number of the demand with attribute $b\in\mathcal{B}$. The demand attribute includes the pick-up location $z^\text{pu}\in\mathcal{Z}$, drop-off location $z^\text{do}\in\mathcal{Z}$, and the demand type $k\in\mathcal{K}$. The demand type set $\mathcal{K}$ consists of serving trip, repositioning, and charging. Note that staying is a type of repositioning (at the same location) as well. The demand attribute is defined as follows:
\begin{align*}
    b=(z^\text{pu},z^\text{do},k).
\end{align*}

\textcolor{black}{It should be noted that, in this model, SoC is the attribute or index of the vehicle state, rather than the state of the Markov decision process. Therefore, the SoC updating is not implemented as a variable; instead, it is updated via the state transition. After determining the pick-up and drop-off locations as well as the demand type, the corresponding trip time $\tau_b$ and energy consumption $e_b$ can be calculated based on the average value of trip request records. The detailed calculation process of the energy consumption can be found in Section \ref{subsec:case study configuration}.} To distinguish diverse demand types, we use superscript to categorize the demand attribute set, i.e. $\mathcal{B}^\text{Trip/Repo/Char}\subset\mathcal{B}$ representing serving trip, repositioning, and charging demand, respectively.
\subsubsection{Decision}
\textcolor{black}{The central operator of the EAMoD system makes various decisions, including serving trip requests for revenue, repositioning to profitable locations, and charging.} As shown in Fig.~\ref{fig:illustration}, the EAMoD dispatch model can be regarded as a network flow problem. Each node (i.e., the battery icon in Fig.~\ref{fig:illustration}) represents a specific vehicle attribute, and the state $R_{t,a}$ is the ``resource" of the node. Different nodes are connected by various types of arcs, indicating various demand types. Building on this network flow basis, the decision is the flow on the arc. Then the constraints of the EAMoD dispatch model can be formulated as follows: 
\begin{align}
    \sum_{\forall b\in\mathcal{B}}&x_{t,a,b}=R_{t,a},\forall a\in\mathcal{A}, \label{cons:flow conservation}\\
    \sum_{\{a|z=z^\text{pu},l>e_b,\forall a\in\mathcal{A}\}}&x_{t,a,b}\leq D_{t,b},\forall b\in\mathcal{B}^\text{Trip}, \label{cons:demand}\\
    &\boldsymbol{x}_t\geq0,\label{cons:non-negative}
\end{align}
\textcolor{black}{where $x_{t,a,b}\in\mathbb{R}$ is the decision variable,} indicating the number of dispatched vehicles with vehicle attribute $a$ to meet the demand with demand attribute $b$. The subscript $a$ of the decision specifies the start node in the network, while the end node can be deterministically obtained (see details in subsection: transition function). \textcolor{black}{Then, the scale of the decision set can be roughly estimated according to $|\boldsymbol{x}|=|\mathcal{Z}|\times(|\mathcal{Z}|-1)\times|\mathcal{L}|$. The practical decision set scale will be slightly smaller due to the elimination rules defined in the 4th row of \textbf{Algorithm \ref{arg:matrix generation}}.} The constraint (\ref{cons:flow conservation}) denotes the flow conservation. As aforementioned, the vehicle state $R_{t,a}$ is the resource in the network framework, thus the flow $x_{t,a,b}$ out of the start node should equal the initial resource. The summation symbol in constraint (\ref{cons:flow conservation}) includes all demand types, meaning all vehicles with attribute $a$ should be dispatched, including serving trips, repositioning, and recharging. For instance in Fig.~\ref{fig:illustration}, two decisions out of start node $a=(1,3)$ connect attribute $a=(1,3)$ and $a=(2,2)$ at time $t=2$, respectively. Then the sum of two flows should equal $R_{t=1,a=(1,3)}$. The constraint (\ref{cons:demand}) denotes that the dispatched EAMoD vehicles for serving trips should not exceed available trip requests. The summation symbol in constraint (\ref{cons:demand}) indicates that only ``available" EAMoD vehicles will be considered. Specifically, given the trip demand $b\in\mathcal{B}^\text{Trip}$, all vehicles at the pick-up location ($z=z^\text{pu}$) and with sufficient SoC ($l>e_b$) are available. Here $z$ and $l$ are vehicle location and SoC, respectively, wherein $z^\text{pu}$ and $e_b$ are pick-up location and required energy consumption for demand attribute $b$, respectively. The inequality (\ref{cons:non-negative}) restrains decisions to be non-negative. The feasible region of the decision variable is represented as $\mathcal{X}_t$.

\subsubsection{Transition Function}In this part, we first introduce how the vehicle attributes are deterministically determined, then we introduce how the future vehicle states are calculated via the transition function. \textcolor{black}{The vehicle attribute, including location and SoC, will be deterministically determined by the specific decision. It should be noted that the state transition depicts the variation of vehicle number at the specific attribute (location and SoC), rather than the variation of the attribute. Nevertheless, the variation of location and SoC can be implicitly reflected via the variation of vehicle number at different attributes. For instance, in Fig.~\ref{fig:illustration}, the vehicles with the attribute $a=(3,3)$ serve trips, and we know the time consumption $\tau_b=2$, energy consumption $e_b=2$ and the drop-off location $z^\text{do}$ is Zone\#1 (according to the exogenous input), then the new attribute must be $a'=(1,1)$ at time $t=3$. During this process, the vehicle number with attribute $a=(3,3)$ reduces, while the vehicle number with attribute $a'=(1,1)$ increases the same value. This deterministic connection relationship between the initial attribute $a$ and the new attribute $a'$ can also be interpreted as network flow theory, that is, the incidence matrix of the network. In other words, we can use an incidence matrix $\boldsymbol{\Delta}^\text{M}$ to depict the connection of the network.} Subsequently, if all movements (i.e. serving trip and repositioning) between zones can be completed in single time point, the new vehicle state $\boldsymbol{R}_{t+1}$ can be calculated via the incidence matrix $\boldsymbol{\Delta^\text{M}}$ and the decision vector $\boldsymbol{x}_t$ as follows:
\begin{align}    \boldsymbol{R}_{t+1}=\boldsymbol{\Delta^\text{M}}\boldsymbol{x}_t.\label{cons:single trans}
\end{align}
\textcolor{black}{Note that the above equation (\ref{cons:single trans}) is equivalent to the redistribution from the state $\boldsymbol{R}_t$ to $\boldsymbol{R}_{t+1}$. However, in this paper, we formulate the problem with higher temporal resolution, thus, some movements (e.g., reposition or serve trips) may take multiple time points to arrive at the destinations. As a result, the new state $\boldsymbol{R}_{t+1}$ is affected by both the past and current decisions; here we call it ``past impact" and ``current impact", respectively. For example, in Fig.~\ref{fig:illustration}, suppose the decision time point is $t=2$, two solid lines from the past ($t=1$) and current ($t=2$) time point, respectively, both connect to the vehicle attribute $a=(1,1)$ at time $t=3$. Consequently, the new state $\boldsymbol{R}_{t=3,a=(1,1)}$ is determined by the past decision from $t=1$ and the current decision. The mathematical formulation can be expressed as:}
\begin{align}    \boldsymbol{R}_{t+1,t+\tau}=\boldsymbol{R}'_{t+1,t+\tau}+\boldsymbol{R}_{t,t+\tau}, \forall\tau=\{1,2,...,\tau^\text{max}\}.\label{cons:state transition}
\end{align}
The symbols in the above equation are added with an additional time subscript. \textcolor{black}{The first time subscript represents the time when the information is received by the system. For instance, $\boldsymbol{R}_{t+1,t+\tau}$ denotes that the future state will be known when the current decision is made, and the system evolves to the next time point.} The second time subscript denotes the time when the vehicles dispatched in the past (or now) arrive at the destination. The symbol $\tau$ denotes the time consumption, and $\tau^\text{max}$ is the maximum time consumption for all movements (including serving trips and repositioning). Subsequently, $\boldsymbol{R}_{t+1,t+\tau}$ denotes the number of vehicles arriving at the destination at time $t+\tau$. The symbol $\boldsymbol{R}'_{t+1,t+\tau}$ is the current impact, and denotes the number of vehicles (dispatched at current time point $t$) arriving at the destination at future time $t+\tau$. The current impact is calculated as follows:
\begin{align}
    \boldsymbol{R}'_{t+1,t+\tau}=\boldsymbol{\Delta}^\text{M}_\tau \boldsymbol{x}_t,\forall\tau=\{1,2,...,\tau^\text{max}\},\label{cons:current impact}
\end{align}
\textcolor{black}{where $\boldsymbol{\Delta}^\text{M}_\tau$ is the incidence matrix mapping the decision $\boldsymbol{x}_t$ to the time $t+\tau$. The symbol $\boldsymbol{R}_{t,t+\tau}$ is the past impact, and denotes the number of vehicles (dispatched at past time points) arriving at the destination at future time $t+\tau$. When the system evolves from the current time $t$ to the next time point $t'=t+1$, the future vehicle state $\boldsymbol{R}_{t+1,t+\tau}$ at time $t$ will become the past impact from the perspective of the time point $t'$. Then the decision $\boldsymbol{x}_{t'}$ will change the past impact $\boldsymbol{R}_{t+1,t+\tau}$ to generate the new vehicle state as the same as we do at time $t$, which is how the system evolves over time. Through multiple state $\boldsymbol{R}_{t+1,t+\tau},\forall\tau$ and mapping matrices $\boldsymbol{\Delta}^\text{M}_\tau,\forall\tau$, we can depict the decisions that span several time points.} 

\textcolor{black}{The incidence matrix $\boldsymbol{\Delta}^\text{M}_\tau$ is the exogenous binary input, which is generated by \textbf{Algorithm \ref{arg:matrix generation}}. Its size is $|\mathcal{A}|\times|\boldsymbol{x}|$. The incidence matrix is inherently related to the decision vector, they share the same columns (length), and each column is related to a specific decision. The incidence matrix is initially set to all zeros. If a decision $x_{t,a,b}$ will dispatch the vehicle to the new vehicle attribute $a'$ with time consumption $\tau_b$, then $\boldsymbol{\Delta}^\text{M}_{\tau_b}(a',d)$ is set to 1, where $d$ is an auxiliary index. Afterwards, the meaning of multiplying the row $a$ of the incidence matrix and the decision vector $\boldsymbol{x}$ is to sum the decision values that send any vehicle attributes to attribute $a$.}
\begin{algorithm}
\caption{Incidence Matrix Generation}
\label{arg:matrix generation}
\begin{algorithmic}[1]
    \STATE Input exogenous demand data, including pick-up $z^\text{pu}$, drop-off location $z^\text{do}$, energy $e$ and time consumption $\tau$
    \STATE Set index $d$=1, $\boldsymbol{\Delta}^\text{M}_\tau=0,\forall\tau=\{1,2,...,\tau^\text{max}\}$
    \FOR{$\forall b\in\mathcal{B}$}
    \FOR{$\{a|z=z^\text{pu}\&l>e_b,\forall a=(z,l)\in\mathcal{A}\}$}
    \STATE Calculate new vehicle attribute $a'=(z^\text{do},l-e_b)$
    \STATE Let $\Delta^\text{M}_{\tau_b}(a',d)=1$
    \STATE Let $d+=1$
    \ENDFOR
    \ENDFOR
\end{algorithmic}
\end{algorithm}

The transition of the demand state is simple, we directly assign the $\boldsymbol{D}_{t+1}$ with the demand data from the exogenous dataset.
\subsubsection{Objective Function}
{Different decisions produce different revenues $r_{t,a,b}$ (vector form $\boldsymbol{r}_t$),} which are given as follows:
\begin{equation}
r_{t,a,b} = 
\begin{cases} 
\lambda^0 + \lambda \cdot w_b, &b\in\mathcal{B}^\text{Trip}\\
-\rho^\text{fast}, &b\in\mathcal{B}^{\text{Char,fast}}, \\
-\rho^\text{slow}, &b\in\mathcal{B}^{\text{Char,slow}}, \\
0,&b\in\mathcal{B}^\text{Repo},
\end{cases}\label{cons:revenue}
\end{equation}
where the revenue from serving trips consists of two parts, including the initial fare $\lambda^0$ and the subsequent distance-related fare $\lambda \cdot w_b$. The revenue from recharging is negative as the cost, while the costs for fast and slow charging are different. In this paper, the reposition cost is set to 0 since the battery consumption has implicitly considered that cost.

Then, the objective of the EAMoD dispatch model is to maximize the total revenue over time as follows:
\begin{align}
    \max_{\boldsymbol{x}\in\mathcal{X}}\{\sum_{t=1}^T(\boldsymbol{r}^\intercal_t\boldsymbol{x}_t|S_t)|S_1\}.\label{obj:bellman}
\end{align}

\subsection{Power System Regulation Model}\label{power model}
The power system regulation model generates a real-time feasible charging range for EAMoD depots according to the short-term load prediction (15 minutes). The urban distribution network is represented by the bus set $\mathcal{N}$ and the line set $\mathcal{U}$, wherein the part of the buses $\mathcal{N}^\text{Depot}$ is connected with the EAMoD depots. The power system regulation model is formulated as follows:
\begin{align}
    \max&\sum_{j\in\mathcal{N}^\text{Depot}}p_j^\text{ev},\label{obj:power range}\\
    \text{s.t.:}~&p^\text{L}_{ij}+p^\text{G}_{j}=\sum_{k\in \mathcal{N}_j}p^\text{L}_{jk}+p_{j}^{\text{D}}+p_{j}^{\text{ev}}, \forall(i,j)\in \mathcal{U}, \label{pbalance}\\
    &q^\text{L}_{ij}+q^\text{G}_{j}=\sum_{k\in \mathcal{N}_j}q^\text{L}_{jk}+q_{j}^{\text{D}}, \forall(i,j)\in \mathcal{U}, \label{qbaalance}\\
    &U_{j}=U_{i}-2(r_{ij}p^\text{L}_{ij}+x_{ij}q^\text{L}_{ij})+(z_{ij})^{2}I_{ij}, \forall(i,j)\in \mathcal{U}, \label{voltagedrop}\\
    &(p^\text{L}_{ij})^{2}+(q^\text{L}_{ij})^{2}\leq U_{i}I_{ij},~\forall(i,j)\in \mathcal{U}, \label{SOCP}\\
    &\underline{U_{j}}\leq U_{j} \leq \overline{U_{j}}, \forall j\in \mathcal{N},\label{Urange}\\    
    &\underline{I_{ij}}\leq I_{ij} \leq \overline{ I_{ij}}, \forall (i,j)\in \mathcal{U}.\label{Irange} 
\end{align}
Note that the time index is omitted in the above model for brevity. The objective function (\ref{obj:power range}) is to maximize the affordable charging power of the given distribution network, where, symbol $p_j^\text{ev}$ is the charging power of the depot connecting to the electrical bus $j$. Constraints (\ref{pbalance})-(\ref{qbaalance}) represent active and reactive power balances, where $p^\text{L}_{ij}$ ($q^\text{L}_{ij}$) is the active (reactive) power flowing from bus $i$ to $j$, $p_{j}^{\text{D}}$ ($q_{j}^{\text{D}}$) is the basic active (reactive) load at bus $j$, and $p_j^\text{G}$ ($ q_j^\text{G}$) is the active (reactive) power injection from the main grid. Constraint (\ref{voltagedrop}) restricts the voltage variation along the line, where $r_{ij}, x_{ij} \text{ and } z_{ij}$ are resistance, reactance, and impedance between the bus $i$ and $j$, respectively. Constraint (\ref{SOCP}) relaxes the original equality to inequality and can be transformed into a second-order cone term. The voltage of the bus and current of the line are bounded by constraints (\ref{Urange}) and (\ref{Irange}).

\textcolor{black}{The above problem is a second-order cone program, which can be efficiently solved by off-the-shelf solvers, thereby ensuring real-time coordination with the EAMoD system. The output charging power range is sent to the EAMoD dispatch system, which then constructs the additional constraints as follows:}
\begin{align}
    \sum_{\{a,b|z=z_j,\forall a\in\mathcal{A},\forall b\in\mathcal{B}^\text{Char}\}}P_bx_{t,a,b}\leq p^\text{ev}_{t,z_j}, \forall z_j\in\mathcal{Z}^\text{Depot},\label{cons:charging power constraints}
\end{align}
where symbol {$P_b$ is the charging piles' rated power}, if $b\in\mathcal{B}^\text{Char,fast}$, $P_b$ is the rated power of fast charging piles, if $b\in\mathcal{B}^\text{Char,slow}$, it is the rated slow charging power; symbol $z_j$ is the transportation zone associated with the power node. 

\textcolor{black}{The proposed power system regulation model estimates the available charging range on the depot nodes by maximizing the sum of the power. However, it does not consider the mutual effect between power nodes during the calculation. Fortunately, this issue does not exist in scenarios where the target nodes connect to different root nodes. Besides, the projection-based feasible region estimation can be readily introduced to avoid this issue, such as reference \cite{tan2019enforcing}.}

\section{Solution Algorithm}\label{solution method}
This section introduces the algorithm to solve the Markov decision process model developed in Section~\ref{robotaix model}. In dynamic programming, the standard Bellman's equation (\ref{obj:bellman}) is divided into several subproblems according to decision points and recursively solved in the following form:
\begin{align}
    V_t(S_t)=\max_{\boldsymbol{x}_t\in\mathcal{X}_t}(\boldsymbol{r}_t^\intercal \boldsymbol{x}_t+\sum_{\tau=1}^{\tau^\text{max}}\{V_{t+\tau}(S_{t+\tau})|S_t,\boldsymbol{x}_t\}),
\end{align}
where $V_t$ is the value function of the subproblem at decision time $t$, and the value functions at future time points $t+\tau$ are affected by the current decision $\boldsymbol{x}_t$. 
\subsection{Piecewise-Linear Value Function Approximation}
Directly solving recursive dynamic programming is computationally expensive due to the curse of dimensionality because of its extremely large state and decision space. To address the computational challenge, we approximate the value function by transforming the unknown and complicated value function $V_t$ into a known and simple function $\tilde{V}_t$, such as a linear or piecewise-linear function. Specifically, we adopt a sample training process to estimate the parameters of the approximated value function $\tilde{V}_t$.

In the EAMoD dispatch problem, it is clear that more vehicles at a location correspond to higher potential revenue, implying a higher value function. However, since trip requests are finite, the revenue will not keep increasing once all requests are satisfied, resulting in diminishing marginal returns. According to this analysis, a piecewise linear function is appropriate to approximate the value function. 

We first introduce the structure of the approximated value function, as shown in Fig.~\ref{fig:PWL}. For each vehicle attribute $a$ at every time point $t$, there is an approximated value function, which is updated over iteration $n$. All symbols with subscript $n$ are related to training iterations. The value for the state with 0 vehicle, $\tilde{V}_{t,n,a}$, must be 0. The piecewise linear function should maintain the concave property, i.e. $v_{t,n,a,m}\geq v_{t,n,a,m+1}, \forall m$, since the concavity is consistent with the diminishing marginal benefit and accurately reflects the real value function's property \cite{godfrey2002adaptivea}. The approximated value function $\tilde{V}_{t,n,a}$ and the corresponding constraints can be represented as follows:
\begin{figure}\color{black}
    \centering
    \includegraphics[scale=0.7]{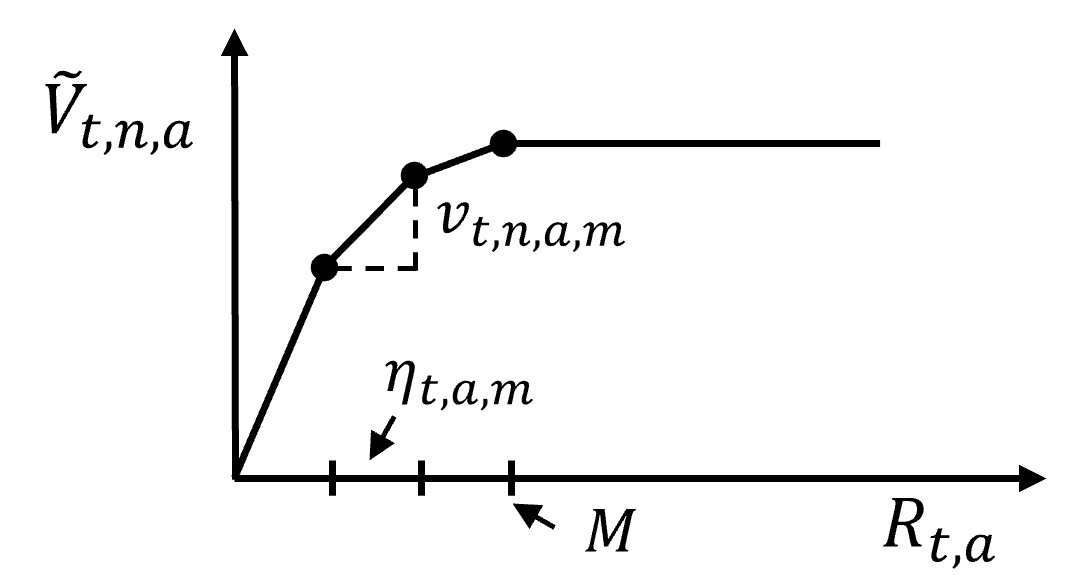}
    \caption{Piecewise-linear value function approximation.}
    \label{fig:PWL}
\end{figure}
\begin{align}
    &\tilde{V}_{t,n,a}=\sum_{m=1}^M v_{t,n,a,m}\eta_{t,a,m},~\forall a\in\mathcal{A},\label{cons:value function}\\
    &\sum_{m=1}^M\eta_{t,a,m}=R_{t,a},~\forall a\in\mathcal{A},\label{cons:eta R}\\
    &0\leq\eta_{t,a,m}\leq1,~\forall a\in\mathcal{A}, \forall m.\label{cons:eta range}
\end{align}
The vehicle state $R_{t,a}$ is divided into $M$ segments with $m$ as the index. \textcolor{black}{The segment number $M$ depicts the segments with positive slopes. If the state $R_{t,a}$ exceeds $M$, the piecewise linear function can still normally depict its value. However, its marginal value (slope) is 0.} Symbol $v_{t,n,a,m}$ is the slope of the $m$th segment, and $\eta_{t,a,m}$ is the length of the $m$th segment. The sum of $\eta_{t,a,m}$ should equal $R_{t,a}$, and the length of each segment is restricted within 0 and 1. The equations (\ref{cons:value function})-(\ref{cons:eta range}) can construct a piecewise-linear value function with respect to the vehicle state, i.e. $\tilde{V}_{t,n,a}(R_{t,a})$. \textcolor{black}{It should be noted that although the introduction of the value function into both the objective function and constraints will not change the linear property of the EAMoD problem because the segment length $\eta_{t,a,m}$ can be the continuous variable if the piecewise-linear function $\tilde{V}_{t,n,a}$ is concave, i.e. $v_{t,n,a,m}\geq v_{t,n,a,m+1}, \forall m$. Besides, the continuous segment length will not lead to issue that $\exists m>R_{t,a}$, such that $0<\eta_{t,a,m}\leq1$. Because if so, considering constraint (\ref{cons:eta R}) and the concavity of the value function, $\Tilde{V}_{t,n,a}$ can always be larger by decreasing $\eta_{t,a,m}$ to zeros and increasing $\eta_{t,a,m'},\exists m'\leq R_{t,a}$, since $v_{t,n,a,m'}\geq v_{t,n,a,m}$.}


Then the overall ADP model can be summarized as follows:
\begin{align}
  \max~&\textcolor{black}{\boldsymbol{r}_t^\intercal \boldsymbol{x}_t+\sum_{\tau=1}^{\tau^\text{max}}\sum_{\forall a\in\mathcal{A}}\sum_{m=1}^M \gamma^{\tau-1}v_{t+\tau,n,a,m}\eta_{t+\tau,a,m},}\label{obj:adp}\\
    \text{s.t.}~&(\ref{cons:flow conservation})-(\ref{cons:non-negative}),(\ref{cons:state transition})-(\ref{cons:current impact}),(\ref{cons:eta R})-(\ref{cons:eta range}).\label{cons:adp cons}
\end{align}
When making decisions, the objective function of the ADP consists of the immediate income (the first term) and the future value (the second term). The immediate income is straightforward, comprising revenue from completing a trip and charging costs as shown in equation (\ref{cons:revenue}). The future value, represented by a piecewise-linear function, reflects the potential value of the location and SoC at a future time. For instance, it is more attractive to accept a trip with a 10-dollar income and 10 potential dollars (because of a new request) at the drop-off location, rather than a trip with a 15-dollar income, but without a trip request at the drop-off location. Furthermore, due to the inherent mechanism of the parameter update (which will be discussed later), the value function at a higher SoC will be higher than that at a lower SoC, leading to natural charging behaviors. \textcolor{black}{Besides, we add a discount factor $\gamma\in[0,1]$ in the value function to reflect the opportunity cost of the multi-period decisions. Specifically, through the discount factor, the future value will gradually decay to depict the phenomenon where the further future represents a higher opportunity cost. As a result, the model will strategically balance the near- and remote-value functions.}

In the ADP, the slope $v_{t,n,a,m}$ is the key parameter to be updated iteratively through samples. The slope denotes the potential value of adding one more vehicle from $m$ at time $t$ and vehicle attribute $a$. In this paper, the dual variable $\pi_{t,a}$ of the flow conservation constraint (\ref{cons:flow conservation}) is adopted to update the slope, since it reflects the marginal value of one additional vehicle. Owing to the property of the linear program, we can obtain the dual variable without additional calculation. The slope at the $n$th iteration is updated according to the following equation:
\begin{align}
    v_{t,n+1,a,m}=(1-\alpha_n)v_{t,n,a,m}+\alpha_n\pi_{t,a},\label{equ:update}
\end{align}
where $\alpha_n$ is the update stepsize at iteration $n$ which decays with the iteration.

The slope update method implies the key feature of the ADP. The dual variable is related to the demand requests of the location. A location with a higher demand level implies that more vehicles can obtain revenue there, leading to a larger updated segment length, and a higher value function. Consequently, the algorithm tends to dispatch more vehicles to such locations until the marginal benefit reduces to 0, which is the reason behind the reposition decision. \textcolor{black}{Furthermore, the algorithm makes charging decisions under the charging and time cost, because the vehicle with a low SoC has far less chance of satisfying trip requests, resulting in a lower associating value function. The gap in value functions leads to the charging behavior. As for the charging selection between the fast and slow chargers, here we take fast charging as an example to analyze. The fast charging can rapidly increase the SoC level, thus increasing more future value, while the current revenue (charging price) will be low. If the high-demand period is about to come, the EAMoD system will naturally prefer the fast charging since the value function will be sufficiently high, even though the charging price is higher than the slow charging.}

The slope update acts only on one segment of the value function, which may disrupt its concavity. To address this issue, we update the associated neighboring segments as well. The concavity maintaining method refers to the book from Powell\cite{powellapproximate}. Let $m'$ denote the updated segment index. If $v_{t,n,a,m'-1}<v_{t,n,a,m'}$, find other segment $m''<m'$, such that $v_{t,n,a,m''}\geq (\sum_{m=m''+1}^{m'}v_{t,n,a,m})/(m'-m'')$. Then we let the slopes between $m''+1$ and $m'$ be their average value. The update method when $v_{t,n,a,m'}<v_{t,n,a,m'+1}$ is symmetrical to the aforementioned process.

The overall training process of the ADP algorithm is summarized in \textbf{Algorithm \ref{arg:adp}}. The training process begins by inputting the exogenous dataset and setting the value function parameters to 0. For each training sample, we initialize the state and solve the EAMoD dispatch model over time. After the problem is solved, the approximated value function is updated according to the obtained dual variable.

\begin{algorithm}
\caption{Approximate dynamic programming training process}
\label{arg:adp}
\begin{algorithmic}[1]
    \STATE Input trip request dataset $\mathcal{N}$
    \STATE Set the initial slope $v_{t,n=1,a,m},\forall t,a,m$ to be 0
    \FOR{$\forall n\in\mathcal{N}$}
    \STATE Reset the initial state $S_1$
    \FOR{$\forall t\in\mathcal{T}$}   
    \STATE Solve the problem (\ref{obj:adp})-(\ref{cons:adp cons})
    \STATE Obtain the dual variable and update the approximated value function based on (\ref{equ:update})
    \STATE Adjust the updated value function to maintain its concavity
    \STATE Update the vehicle state according to (\ref{cons:state transition})  
    \ENDFOR
    \ENDFOR
\end{algorithmic}
\end{algorithm}

\subsection{Approximate Dynamic Programming with Model Predictive Control}\label{subsec:adpmpc}

\textcolor{black}{Although the basic ADP algorithm can basically complete the decision-making, the limitation remains, especially in problems with an accurate charger accessibility modeling. As aforementioned, all decisions are based on the immediate revenue and future value function. However, at some time points, the future value function of the charging may exceed that of some trip-serving decisions, especially when the drop-off locations have low demand. This confusion between charging and serving trip requests can lead to behaviors where the vehicles first charge to a high SoC, and then stay at the depot. Consequently, the utilization rate of EAMoD fleets is significantly reduced, and a large number of trip requests cannot be met. We notice that the limitation mainly consists in that the basic ADP algorithm makes decisions solely based on the value function, while it cannot clearly distinguish the value of the SoC and location. Therefore, to compensate for the limitation of the value function, we combine the basic ADP with the MPC. With the additional information input from only one more horizon, the proposed algorithm can make better decisions than the basic ADP algorithm, while still maintaining good computational efficiency compared to the long-horizon MPC algorithm.} The model in the proposed algorithm is formulated as follows:

\begin{align}
\max~&\begin{aligned}[t]
    &\sum_{h=0}^H\boldsymbol{r}_{t+h}^\intercal \boldsymbol{x}_{t+h}\\
    &\textcolor{black}{+\sum_{\tau=1}^{\tau^\text{max}}\sum_{\forall a}\sum_{m=1}^M \gamma^{\tau-1}v_{t+\tau+H,n,a,m}\eta_{t+\tau+H,a,m},}
\end{aligned}    \label{obj:adpMPC}\\
    &\sum_{\forall b\in\mathcal{B}}x_{t+h,a,b}=R_{t+h,a},\forall a, h,\label{cons:re flow conservation}\\    
    &\sum_{\{a|z=z^\text{pu},l>e_b,\forall a\}}x_{t+h,a,b}\leq D_{t+h,b},\forall h,b\in\mathcal{B}^\text{Trip},\label{cons:re demand}\\   
    &\boldsymbol{x}_{t+h}\geq0,\forall h, \label{cons: re non-negative}\\&\boldsymbol{R}'_{t+h+1,t+h+\tau}=\boldsymbol{\Delta}^\text{M}_{\tau}\boldsymbol{x}_{t+h},\forall \tau,h,\label{cons:re corrent impact}\\    &\boldsymbol{R}_{t+h+1,t+h+\tau}=\boldsymbol{R}'_{t+h+1,t+h+\tau}+\boldsymbol{R}_{t,t+\tau},\forall \tau,h,\label{cons: re state transition}\\
    &\sum_{m=1}^M\eta_{t+\tau+H,a,m}=R_{t+\tau+H,a},\forall a,\tau,\label{cons: re eta R}\\
    &0\leq\eta_{t+\tau+H,a,m}\leq1,\forall a,\tau,m.\label{cons: re eta range}
\end{align}
The reformulated model involves an additional subscript $h={0,1,...,H}$. It should be noted that $R_{t,a}$ is equivalent to $R_{t,t,a}$. In this study, we only need one more horizon to remove algorithmic confusion, i.e., $H=1$. Note that though the transition function in the ADP subproblem is changed as in (\ref{cons:re corrent impact})-(\ref{cons: re state transition}), the state transition is only updated by $\boldsymbol{x}_t$, instead of $\boldsymbol{x}_{t+h},\forall h>0$.

\section{Experiments}\label{result}
\subsection{Case Study Configuration}\label{subsec:case study configuration}
\begin{figure}\color{black}
    \centering
    \includegraphics[scale=0.5]{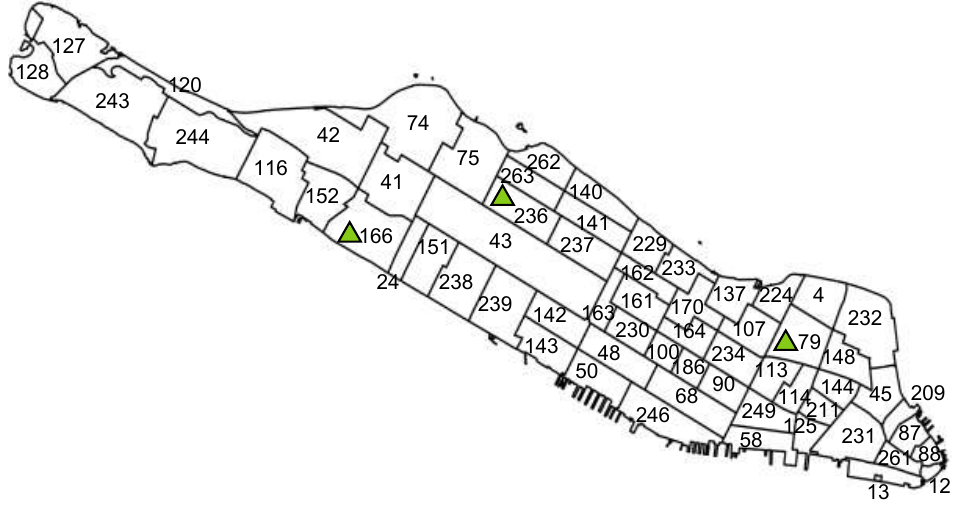}
    \caption{Manhattan with 61 taxi zones\cite{tlc}.}
    \label{fig:manhattan}
\end{figure}

\begin{figure}\color{black}
    \centering
    \includegraphics[scale=0.55]{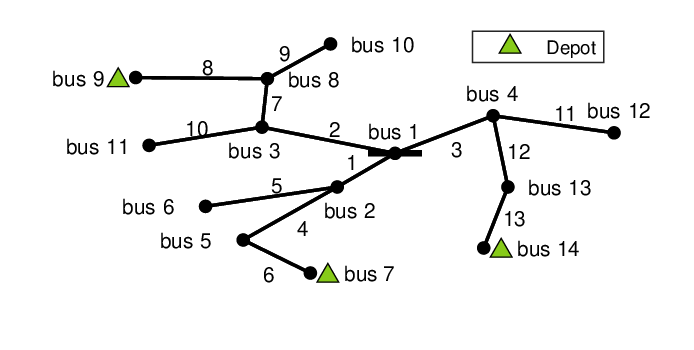}
    \caption{A 14-node distribution network\cite{zhang2018pev}.}
    \label{fig:power}
\end{figure}

We validate the proposed real-time coordination framework of the EAMoD dispatch model and power system regulation model in Manhattan city (Fig.~\ref{fig:manhattan}) and the 14-node distribution network (Fig.~\ref{fig:power}). The number on each block of the Manhattan network is the taxi zone index. The coupled system operates from 5:00 AM to 3:00 AM of the following day, wherein the EAMoD dispatch system makes decisions every 5 minutes, and the power network schedules every 15 minutes. Vehicles outside the depot at the end of the operation time automatically return to depots for charging. \textcolor{black}{The trip data is sourced from New York yellow taxi trip records for 2023 \cite{tlc}. In order to focus on Manhattan, we filter the raw data to save orders with Manhattan as the origin and destination. Besides, the trips within the same transportation zone are not considered. Since the transportation zone division is based on the neighborhood, every zone can reflect the specific travel demand feature, thus, the trip within the same zone only accounts for a small portion (5.5\%). Then, we filter the data of all public holidays to maintain similar trip characteristics. After data preparation, there are 260 samples available for training and testing. The daily order numbers range from 55,000 to 57,000, with 56324 average orders.} We assume that the trips with the same pick-up and drop-off locations are the same. The travel distance and time consumption are all assigned by the average value of the same trips. The different types of trip requests are aggregated into the period of 5 minutes. The EAMoD system has 600 autonomous electric vehicles to meet the trip request, with three EAMoD depots in the model. The connection relationship between transportation zones and power buses is presented in Table~\ref{tab:network connection}. Each depot is equipped with sufficient fast (60kW) and slow (7.5kW) charging piles. The trip fare follows the standard metered fare rule of the New York yellow taxi, where $\lambda^0$ is \$3, and the distance-related fare $\lambda$ is \$2.2 per kilometer\cite{taxi}. The charging prices $\rho$ for fast and slow charging are 1 \$/kWh and 0.2 \$/kWh, respectively, referring to the New York public charging station website\cite{chargefinder}. The battery capacities of EAMoD fleets are assumed to be the same, with 40 kWh, and are divided into 40 levels. Considering the battery degradation from over-charge and discharge, the batteries are assumed to work in the range from 0 to 80\%. Therefore, each SoC denotes a discretization of 32 kWh (i.e. $0.8\times40$). Referring to the Zeeker-X, which is the vehicle type of Waymo, the EV consumption is between 16.4–18.3 kWh/100 km. Accounting for the additional power from radar and computing units, we assume the EV consumption to be 20 kWh/100 km. \textcolor{black}{With given battery capacity and the energy efficiency, we can calculate the equivalent driving distance of each SoC $l$. Then we can round the all trip distance according to the equivalent driving distance, and transform the trip distance into the equivalent SoC consumption. The SoC calculation for fast and slow charging is similar to that for trip, we can calculate the charged energy according to the rated power and the time slot, and then transform it to the SoC form. The initial state of the vehicles for all samples and tests is the same, where the vehicles are evenly located in 3 depots (200 vehicles for each depot), and the average SoC is set to 10.} 

\textcolor{black}{The modeling tool used in this paper is YALMIP \cite{lofberg2004yalmip}. The subproblem of the EAMoD dispatch problem at each decision period, and the power system regulation model are both solved via linear program solver and second-order cone solver of Gurobi \cite{gurobi}, respectively. All coding and solving works are completed on a laptop with i5-13500HX CPU.}

\begin{table}[]
\caption{Network Configuration}
\centering
\begin{tabular}{cccc}
\hline
Depot   & \multicolumn{1}{c}{Depot\#1} & \multicolumn{1}{c}{Depot\#2} & \multicolumn{1}{c}{Depot\#3} \\ \hline
Transportation zone & 79& 166& 236\\
Power bus& 14 & 9 &7 \\ \hline
\end{tabular}\label{tab:network connection}
\end{table}

\subsection{Algorithm Performance Analysis}
\textcolor{black}{We compare the proposed algorithm with three benchmarks: 1) the basic ADP algorithm, 2) the MPC algorithm, and 3) an ideal algorithm that assumes perfect forecast information.} \textcolor{black}{The parameter settings are the same for both the proposed algorithm and the basic ADP algorithm, with the update stepsize $\alpha$ decaying from 0.1 to 0.01, and maximum segment $M$ 20.} The horizon of the proposed algorithm is 1, while the horizon of the MPC algorithm is set to 10 (i.e., 50 minutes accurate trip request prediction). The ideal algorithm with perfect forecast information is equivalent to the MPC algorithm with $H=263$, where all future information is assumed to be known. However, the ideal strategy with perfect information has 264-fold more constraints and variables compared to the model with a single time point, leading to an intractable problem. As the EAMoD dispatch model is a large-scale sparse problem, the commercial solver will generally first use the barrier algorithm to rapidly find an interior point, whose objective value is very close to the optimal one, and then the solver will push the interior point to the accurate basic solution. \textcolor{black}{Therefore, we use the objective value of the interior point as the ideal upper bound, whereas the accurate solution of the decision set is not found.}

\begin{figure}
    \centering
    \includegraphics[width=0.8\linewidth]{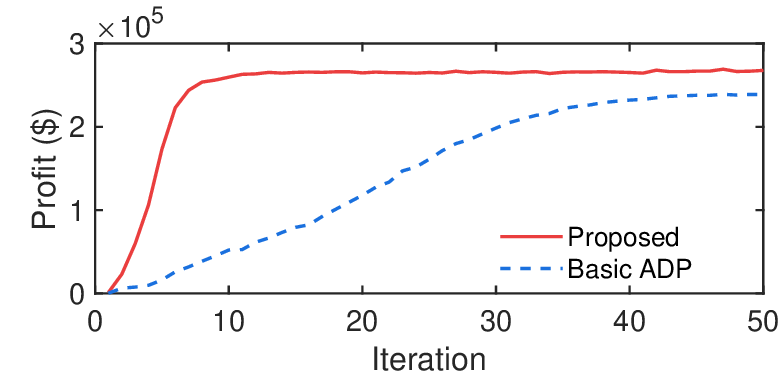}
    \caption{Training process of the proposed and the basic ADP algorithms.}
    \label{fig:train}
\end{figure}

We adopt 50 samples to train the proposed algorithm and the basic ADP algorithm, and 10 samples to test the algorithms. The training processes of the proposed algorithm and the basic ADP algorithm are illustrated in Fig.~\ref{fig:train}. \textcolor{black}{Although both the basic ADP and the proposed algorithm cannot find the global optimum due to the uncertain environment, the proposed algorithm shows significant superiority over the basic ADP algorithm in terms of optimal value and convergence speed. Owing to the additional future information with one more horizon, the proposed algorithm can naturally find a better solution than the basic ADP algorithm. Specifically, the basic ADP algorithm makes decisions solely based on the value function, but it fails to clearly distinguish the value difference between SoC and location (see details in Section \ref{subsec:adpmpc}). Noticing the reason for this limitation, the proposed algorithm introduces an additional horizon to support the value function. The fast convergence is also attributed to the excellent solution. Since the parameter update of the ADP algorithm fully depends on the dual variable, a better solution can better reflect the value of the location or the SoC, which, in turn, outputs better dual variables to update the algorithm. Therefore, the proposed algorithm can find a good solution in the initial iterations and keep exploiting for a better one, leading to faster convergence.}

\begin{figure}\color{black}
    \centering
    \includegraphics[scale=0.45]{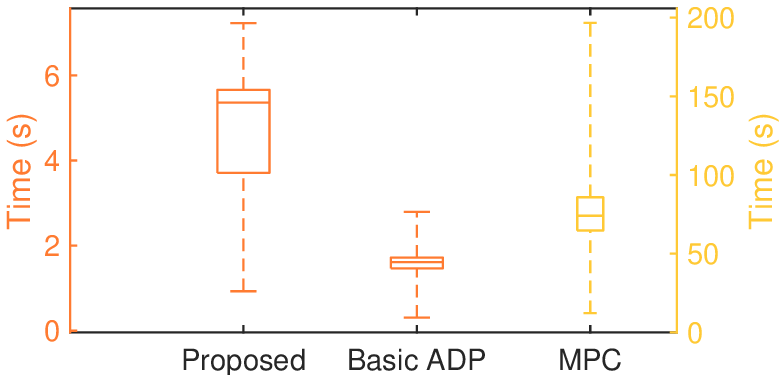}
    \caption{Calculation time of one decision.}
    \label{fig:time cost}
\end{figure}

Then we verify the algorithms in 10 test samples (10 days). \textcolor{black}{The daily average profit, quality of service, daily average profit per trip, and standard deviation of daily profit in 10 test days are illustrated in the Table~\ref{tab:algorithm results}. The daily average profit by the proposed algorithm is 12.1\% and 5.1\% higher than the basic ADP algorithm and the MPC algorithm, respectively. It is only 7.9\% less than the ideal solution, indicating the proposed algorithm can find a near-optimal solution.} Furthermore, the performance of the proposed algorithm shows better stability of daily average profit across the test samples, as indicated by the standard deviation results. The quality of service reflects the ratio of satisfied trips to total trips. The proposed algorithm can meet 92.3\% trips, which is higher than the other two algorithms. Apart from higher daily profit, stability, and quality of service, the proposed algorithm can also achieve higher average profit per trip, meaning it has better dispatch efficiency. Beyond the optimality and stability of the algorithm, computational efficiency is also vital for the real-time operation algorithm. We use the box diagram to present the calculation time of one decision, as shown in Fig.~\ref{fig:time cost}. Note that the box diagram summarizes the statistical results of the calculation times of 2640 decisions in total 10 test days. The time cost of the three algorithms shows that the MPC algorithm costs far more time to make one decision compared to the two ADP-based algorithms. This is because the MPC algorithm ($H=10$) has 11 times more constraints and variables than the ADP algorithm. While most of these constraints and variables can be relaxed during time points with low trip demand, the MPC algorithm remains efficient at these decision points. However, the complexity of the model increases dramatically after constraints and variables are activated. As a result, the MPC algorithm may take nearly 200 seconds to complete one decision, while the next decision point is about to come.

\begin{figure*} \color{black}
\vspace{-5mm}
	\centering
    \subfloat[\label{fig:unbound_behavior}]{
		\includegraphics[scale=0.5]{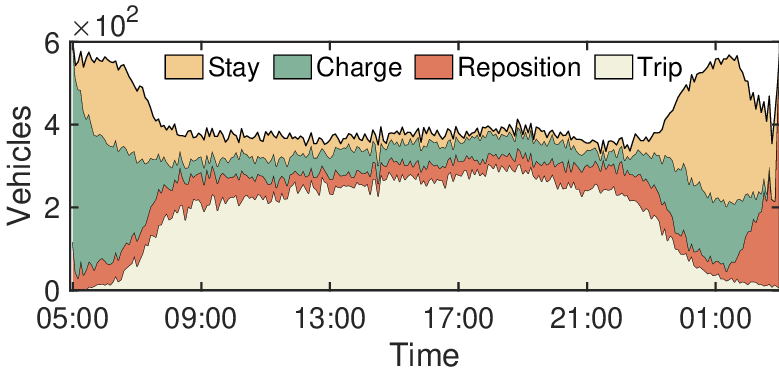}}
        \subfloat[\label{fig:unbound_state}]{
		\includegraphics[scale=0.5]{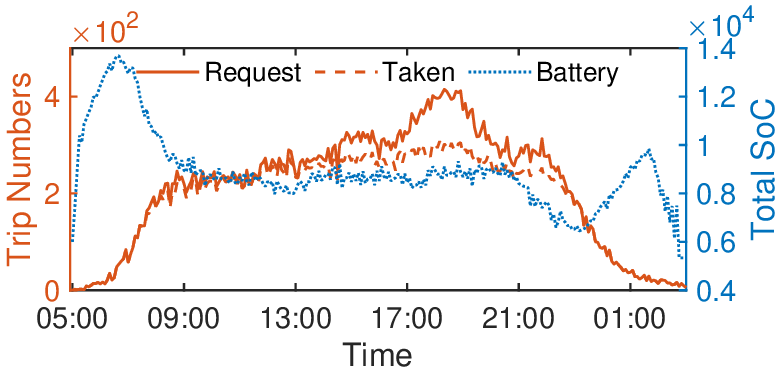}}
        \\[-12pt]
	\subfloat[\label{fig:Nompc_behavior}]{
		\includegraphics[scale=0.5]{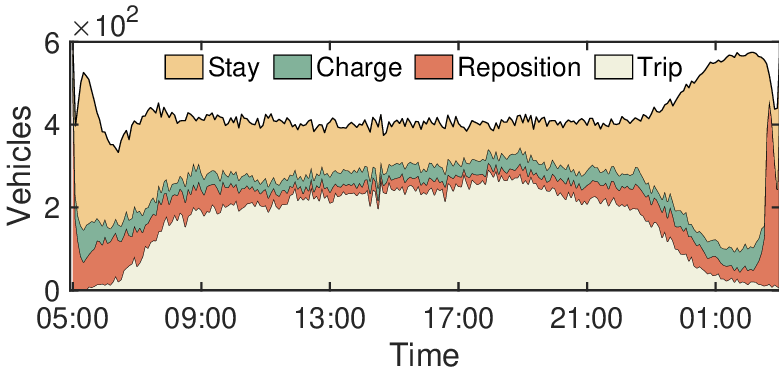}}
        \subfloat[\label{fig:Nompc_state}]{
		\includegraphics[scale=0.5]{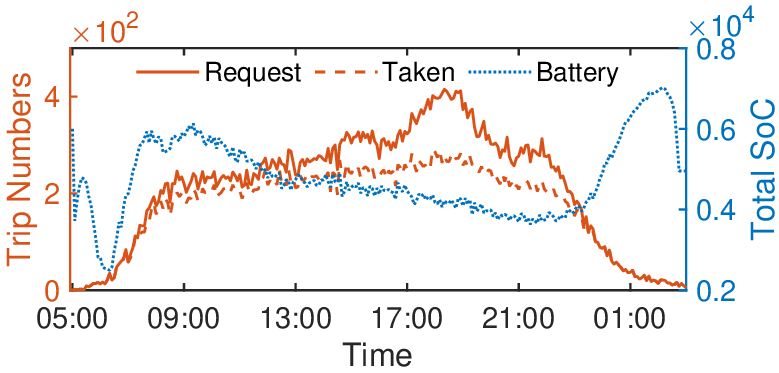} }
        \\[-12pt]
	\subfloat[\label{fig:MPC10_behavior}]{
		\includegraphics[scale=0.5]{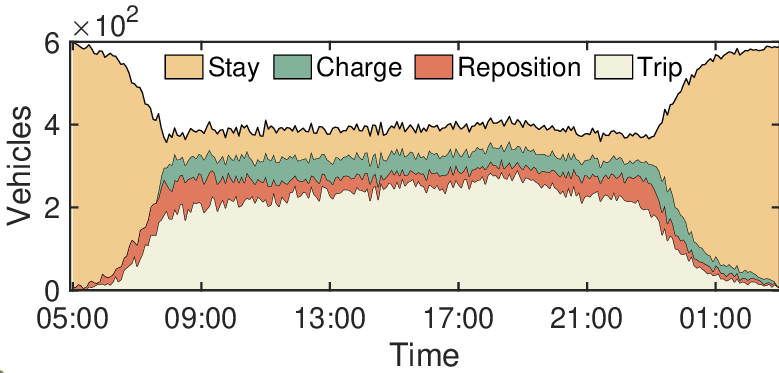}}
	\subfloat[\label{fig:MPC10_state}]{
		\includegraphics[scale=0.5]{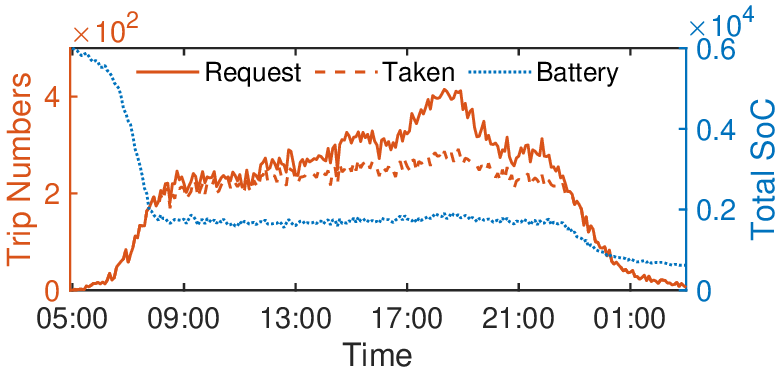}}
	\caption{Behaviors (the first column) and states (the second column) of EAMoD fleets under different algorithms. 
    (a) and (b): the proposed algorithm;
    (c) and (d): basic ADP algorithm;  (e) and (f): MPC algorithm.}
	\label{fig:beh&state} 
\end{figure*}

\begin{table}[]\color{black}
\caption{Algorithm Performance in Test Set}
\centering
\begin{tabular}{ccccc}
\hline
Algorithm& \multicolumn{1}{c}{Proposed} & \multicolumn{1}{c}{Basic ADP} & \multicolumn{1}{c}{MPC} & Ideal     \\ \hline
Average profit (k\$) & 268 & 239  & 255 & 291 \\
Standard deviation (k\$) & 0.71 & 1.15  & 1.62 & 0.44    \\
Quality of service (\%) & 92.3 & 86.4 & 89.5 & -\\
Average profit per trip (\$) & 5.40 & 5.27 & 5.38 & -\\\hline
\end{tabular}\label{tab:algorithm results}
\end{table}

\textcolor{black}{We further analyze the behaviors and states of the EAMoD fleets under the proposed and the benchmark algorithms to better understand the superiority of the proposed one, as shown in Fig.~\ref{fig:beh&state}.} The figure is based on one of the results in the test set. In the figure, the plots in the first column represent the autonomous electric vehicles' behaviors, while the plots in the second column represent the vehicles' states. In the legend of state plots, ``request" refers to the trip demand, ``taken" denotes the satisfied demand, and the ``battery" represents the sum of all autonomous electric vehicles' SoC. We first focus on the behaviors of the basic ADP algorithm (Fig.~\ref{fig:Nompc_behavior}) and the proposed algorithm (Fig.~\ref{fig:unbound_behavior}). It is clear that the stay decisions are more frequently made in the ADP algorithm than in the proposed algorithm. Since the ADP algorithm makes decisions according to the tradeoff between future value and current revenue, the additional charging decision can confuse this mechanism when the value function at a high SoC is high enough, as we analyzed in subsection~\ref{subsec:adpmpc}. As a result, a large number of vehicles charge first and then stay at the depot to benefit from the high future value, which underscores the importance of introducing an additional horizon to balance the complicated value function between decisions. The reason why the MPC algorithm, though having far more prediction information, still shows worse performance than the proposed algorithm can be answered by Fig.~\ref{fig:beh&state}. The battery level curve in Fig.~\ref{fig:MPC10_state} shows an almost monotonic decrease throughout the operation period due to the relative myopia property of the MPC algorithm. The MPC algorithm only finds the global optimum within the optimized horizon, thus, it tends to drain all of the battery energy. As a result, the available vehicles (SoC is greater than 1) gradually decrease, leading to a local optimum when considering the full operation period. Furthermore, the sum of the SoC at the end is 606 (that is, the average SoC is 1.01), which means that most vehicles use out of battery energy. 

\subsection{Impact Analysis of Power Network Constraints}
In this section, we analyze the necessity of considering power network constraints and the impact on both the EAMoD system and the power network. When considering the power network constraints, we add the dispatch model's constraint set (\ref{cons:re flow conservation})-(\ref{cons: re eta range}), along with additional charging regulation constraints (\ref{cons:charging power constraints}). After obtaining the behavior results of two cases (i.e., whether considering the charging regulation constraint), we transform the charging behaviors into the power demand and solve the optimal power flow to check the operation state of the power network. Note that the optimal power flow may be infeasible when the charging power exceeds the safety range. In such scenarios, we relax the voltage and current constraints in the optimal power flow model and assume that the power network can still provide charging power when the voltage constraint is activated. This trick is used to showcase the impact of unregulated charging on the power grid.

\begin{figure}
\centering
    \subfloat[\label{fig:voltage}]{
		\includegraphics[scale=0.45]{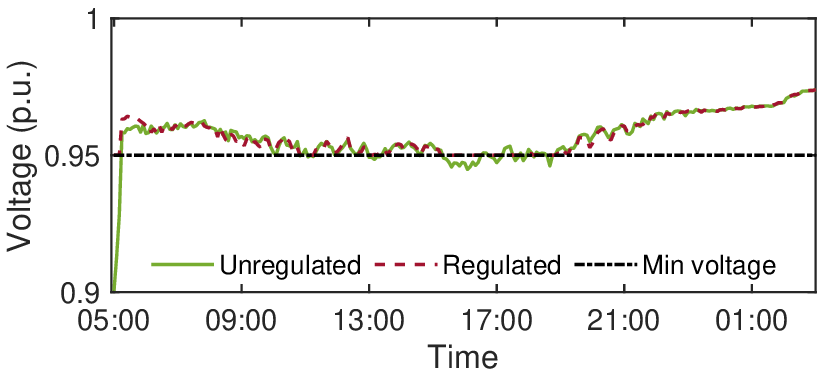}}
        \\[-2pt]
    \subfloat[\label{fig:charge curve}]{
        \hspace*{4pt}
		\includegraphics[scale=0.45]{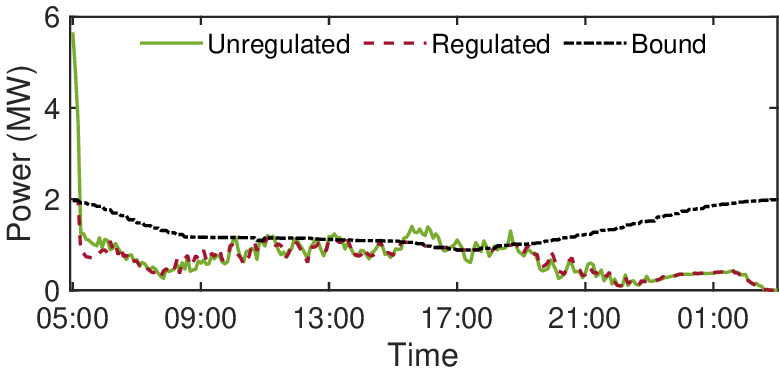}} 
        \\[-2pt]
    \subfloat[\label{fig:fast slow}]{
	\includegraphics[scale=0.45]{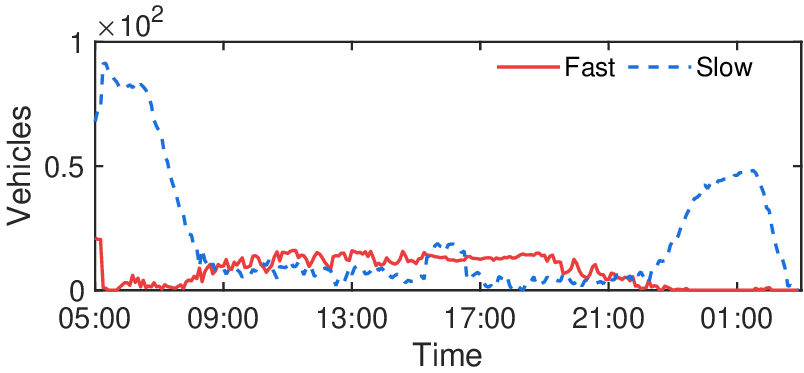}}
    \caption{The operation state of the the power bus 9. (a) node voltage curve; (b) node charging power curve; (c) usage of the fast and slow charging piles with power network regulation.}
    \label{fig:voltage&charge}
\end{figure}

\begin{figure}\color{black}
    \centering
    \includegraphics[scale=0.6]{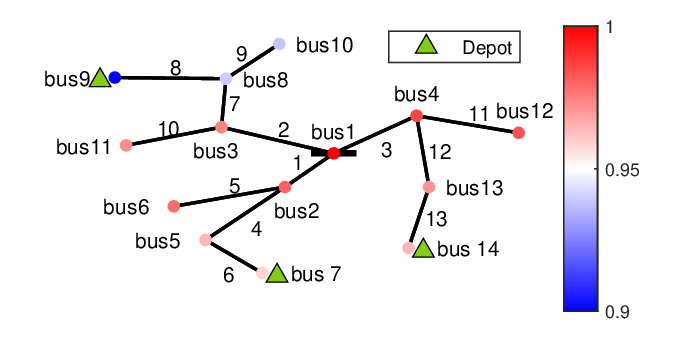}
    \caption{Network nodal voltages at 5:00 in unregulated scenario.}
    \label{fig:voltage_map}
\end{figure}

The experiment results show that the undervoltage occurs at the power node 9. The voltage, the charging power curve, and the usage of the fast and slow charging piles with charging regulation constraints at the power bus 9 over the operation period are shown in Fig.~\ref{fig:voltage&charge}. The ``unregulated" denotes the case without charging regulation constraint, while the ``regulated", on the contrary. The voltage lower bound in this study is set to 0.95 p.u. The ``bound" in Fig.~\ref{fig:charge curve} represents the charging range set by the power network. The impact of the power network regulation on the EAMoD system is summarized in Table~\ref{tab:overall cost}, including average profit in 10 test samples, quality of service, and average profit per trip. In Fig.~\ref{fig:voltage}, abnormal voltage mainly occurs at two periods. At the beginning of the operation day (i.e., from 5:00 to 5:10), trip demand is very low, thus the EAMoD system takes advantage of the sufficient time and dispatches fleets to charge simultaneously. As a result, the charging power at the depot exceeds the limit, leading to severe undervoltage down to 0.9 p.u. \textcolor{black}{Besides, we present the network nodal voltages at 5:00 in Fig.~\ref{fig:voltage_map}. The result shows that the high charging demand of bus 9 not only results in the severe undervoltage at the local node, but it also spreads the voltage drop below 0.95 p.u. at neighbor nodes (bus 8 and 10).} Another abnormal voltage period is from 15:30 to 16:30. The power demand increases gradually over time, thus accommodating less charging power. 

In the ``regulated" case, the charging power is always safely below the safety bound due to the real-time charging regulation constraints. Although the charging power is restricted below the bound, the EAMoD system can still replace the fast charging piles with the slow piles to refill part battery energies as shown in Fig.~\ref{fig:fast slow}. \textcolor{black}{Table~\ref{tab:overall cost} presents the impact of the power network regulation on the EAMoD system in terms of average profit over test samples, quality of service, and average profit per trip. Owing to the flexible charging choice, the average profit and quality of service losses of the EAMoD system by the power network regulation are negligible (only 0.37\% and 0.11\% losses, respectively). Meanwhile, the average profit per trip remains the same, indicating the ``regulation" does not affect the dispatch efficiency.} Apart from the switch of fast and slow charging piles during the charging regulation constraints being activated, the combination of fast and slow charging piles also brings more flexible charging decisions during other operation periods. When the trip requests are low and the time is sufficient during valley hours, the EAMoD system chooses cheap slow-charging piles. During the relatively high-demand hours, the EAMoD system chooses fast-charging piles to rapidly refill battery energies, and then take trip requests.

\begin{table}[]
\centering
\caption{\textcolor{black}{Impact of Power Network Regulation on EAMoD System}}
\begin{tabular}{ccc}
\hline
              & Unregulated & Regulated \\ \hline
Average profit (k\$)    & 268     & 267   \\
Quality of service (\%) & 92.3& 92.2\\
Average profit per trip (\$) & 5.40 &  5.40\\\hline
\end{tabular}\label{tab:overall cost}
\end{table}

\section{Conclusion}\label{conclusion}
This paper proposes a real-time coordination framework integrating the EAMoD system and the power network. The proposed model, based on the Markov decision process, comprehensively accounts for the temporal-spatial characteristics of the autonomous electric vehicles, including serving trips, repositioning, and charging, as well as the operation safety of the power network. To handle the limited decision quality attributed to the complex decision space, we propose a combined ADP and MPC algorithm. 

We then implement the numerical experiments in the Manhattan transportation network and the 14-node power network to verify the performance of the proposed algorithm and evaluate the impact of the system coordination on the interdependent system. In terms of algorithm performance, the proposed algorithm can converge far faster and achieve better profit than the basic ADP. Compared to the MPC with 10 horizons, the proposed algorithm with only 1 horizon can still obtain a 5.1\% better and more stable solution, while spending far shorter CPU time for one decision. \textcolor{black}{Apart from the fundamental comparison of the computational performance, we conduct an in-depth algorithm comparison according to the vehicle behaviors, wherein the basic ADP algorithm assigns a lot of vehicles to stay, leading to poor operation efficiency, while the MPC algorithm exhibits a myopic charging strategy.} In the impact analysis of the power network constraints, we find that the coordination of two systems can effectively prevent undervoltage, while the profit of the EAMoD system remains almost unaffected due to the flexible charging choice between fast and slow charging piles.

\textcolor{black}{Due to the good model generality, we will expand our model to consider both the vehicle-to-grid and renewable energy accommodation in our future work. However, the proposed model cannot achieve accurate vehicle-level dispatch due to its flow-based formulation. Therefore, a hierarchical dispatch framework can be introduced in the future, where the proposed model provides the regional dispatch, then a vehicle-routing model provides the locational dispatch.}

\bibliographystyle{ieeetr}
\bibliography{AEV}

\newpage

 




\vfill

\end{document}